\documentclass{article}

\usepackage{arxiv}

\usepackage[utf8]{inputenc} 
\usepackage[T1]{fontenc}    
\usepackage{hyperref}       
\usepackage{url}            
\usepackage{amsmath,amsthm,amsfonts,amssymb,cancel}       
\usepackage{mathtools}
\usepackage{lipsum}
\usepackage{graphicx}
\usepackage{etoolbox}
\graphicspath{ {./images/} }

\newcommand{\pt}[1]{\left(#1\right)}
\newcommand{\ct}[1]{\left[#1\right]}
\newcommand{\bp}[3]{\int_{0}^{1} \ifstrequal{#1}{0}{}{\theta^{#1}} \ifstrequal{#2}{0}{}{\left(1-\theta\right)^{#2}}\ifstrempty{#3}{}{#3}d\theta}
\newcommand{\bm}[2]{\int_{0}^{\theta}\ifstrequal{#1}{0}{}{q^{#1}} \ifstrequal{#2}{0}{} {\left(1-q\right)^{#2}}dq}
\newcommand{\tsum}{\pt{1+\sum_{i=1}^{m}{n_i}}}
\newcommand{\ssum}[1]{\sum_{i=1}^{m}{#1}}
\newcommand{\jsum}[1]{\sum_{j=1}^{m}{#1}}

\newcommand{\gtmat}[6]{
	\bigl(\begin{smallmatrix}
		#1 & \ldots & #3 & \ldots & #5 \\
		#2 & \ldots & #4 & \ldots & #6
	\end{smallmatrix}\bigr)
}

\newcommand{\tmatt}[4]{
	\begin{pmatrix}
		#1 & #3 \\
		#2 & #4
	\end{pmatrix}
}

\title{An exact, unconditional, nuisance-agnostic test for contingency tables}

\author{
 Miguel Araujo-Voces \\
  Departamento de Bioqu\'imica y Biolog\'ia Molecular\\
  Universidad de Oviedo, Oviedo, Spain\\
  \texttt{araujomiguel@uniovi.es} \\
   \And
 V\'ictor Quesada \\
  Departamento de Bioqu\'imica y Biolog\'ia Molecular\\
  Universidad de Oviedo, Oviedo, Spain \\
  CIBERONC, Madrid, Spain \\
  \texttt{quesadavictor@uniovi.es} \\
}

\begin{document}
\maketitle
\begin{abstract}
Exact tests greatly improve the analysis of contingency tables when marginals are low. For instance, researchers often use Fisher's exact test, which is conditional, or Barnard's test, which is unconditional but needs to deal with a nuisance parameter. Here, we describe the m-test, an exact, unconditional test for the study of \emph{d x m} binomial contingency tables. When comparing binomial trials, the m-test is related to Barnard's exact test. However, the nuisance parameter is integrated over all its possible values, instead of maximized or otherwise estimated. According to Monte Carlo simulations, this strategy yields a higher statistical power than other exact tests. We also provide a package to perform the m-test in \texttt{R}.
\end{abstract}


\section*{Introduction}
Numerous experimental designs involve repeatedly measuring an outcome in multiple groups to determine whether the probability of those outcomes is the same in all groups. Such experiments yield results that can be expressed in the form of a contingency table. Predictably, researchers have developed multiple statistical tests to extract conclusions from these tables \cite{recommended}. The radical improvement of our computational capacity in the last decades has renewed the interest for exact tests that take into account the combinatorial nature of these experiments.

The most widely used exact test to study \emph{2 x 2} contingency tables is Fisher's test \cite{Fisher}. The model for this test is conditional to both marginals for the table being fixed. The natural way for this condition to be true is for the experiment to exhaust all the elements in each group. For instance, in the \textit{lady tasting tea} experiment described by Fisher, all cups prepared are tasted \cite{Design}. Relaxing this constraint leads to unconditional tests, which, in exchange for this advantage, need to deal with a \textit{nuisance parameter}. Since the description of this problem and possible solutions by Barnard \cite{Barnard}, multiple treatments have been developed and tested for this problem \cite{fview}.

Here, we present the m-test, an unconditional binomial exact test that integrates all possible values of the nuisance parameter. With enough computing power, the test is easy to apply to general \emph{d x m} contingency tables. We have also developed a freely available \texttt{R} package to perform the m-test.

\section*{Two-sided \emph{2 x m} test}

The m-test assumes a model similar to that of Barnard's test. 
We compare multiple {\em experiments} ($E$), where a dichotomic outcome ({\em success} or {\em failure}) is repeatedly measured $n$ times under the assumption that the probability of success $\theta$ is always the same (Bernouilli trials). The result of $E$ can then be expressed as the number of successes ($s$) and failures ($f$). If $m$ independent experiments are performed on the same system with $n_i=s_i+f_i$ trials per experiment, the probability $P$ of a set of results is:

\begin{equation}\label{eq1}
P \gtmat{s_1}{f_1}{s_j}{f_j}{s_m}{f_m} = \binom{n_1}{s_1}\ldots\binom{n_j}{s_j}\ldots\binom{n_m}{s_m}\theta^{\ssum{s_i}}\pt{1-\theta}^{\ssum{f_i}}
\end{equation}

In this equation, $\theta$ is called the nuisance parameter, as its value is unknown in the general case. 
Taking this limitation into account, the m-test considers all its possible values by integrating \eqref{eq1} over the probability space:

\begin{equation}\label{eqint}
P \gtmat{s_1}{f_1}{s_j}{f_j}{s_m}{f_m} =\ct{\prod_{i=1}^{m}\binom{n_i}{s_i}}\int_{0}^{1} \theta^{\ssum{s_i}}\left(1-\theta\right)^{\ssum{f_i}}d\theta
\end{equation}

Although \eqref{eqint} can be solved by using the Euler integral of the first kind, it is more convenient to observe some properties, as outlined in Proofs 1 and 2 of the Supplementary Material:

\begin{equation}\label{eqsyst}
\begin{split} 
P \gtmat{0}{n_1}{0}{n_j}{0}{n_m} &= \frac{1}{\tsum} \\
P \gtmat{s_1}{f_1}{s_j+1}{f_j - 1}{s_m}{f_m} &= \frac{f_j\pt{1+\ssum{s_i}}}{\pt{s_j+1}\ssum{f_i}} P \gtmat{s_1}{f_1}{s_j}{f_j}{s_m}{f_m}
\end{split}
\end{equation}

The equations in \eqref{eqsyst} can be used to build an iterative algorithm to compute every possible result with fixed $n_i$ values (Figure \ref{fig:fig1}). In turn, this allows the fast calculation of p-values with fixed column marginals. 
Notably, these equations can be applied to \emph{d x m} contingency tables with a minor tweak, as shown in Proofs 3 and 4 of the Supplementary Material.

\section*{One-sided \emph{2 x 2} test}

Similar to the two-sided test, in a \emph{2 x 2} contingency table we can integrate the nuisance parameters under a null hypothesis stating that the probability of success in the first experiment is higher than in the second experiment:

\begin{equation} \label{os}
P^{-}\tmatt{s_1}{f_1}{s_2}{f_2} = \binom{n_1}{s_1} \binom{n_2}{s_2}\bp{s_1}{f_1}{\bm{s_2}{f_2}}
\end{equation}

In \eqref{os}, the second nuisance parameter $q$ \textemdash  the probability of success in the second experiment \textemdash  is integrated over all the values lower than $\theta$. Similar to the two-sided test, and as shown in Proofs 5-7 of the Supplementary Material,

\begin{equation} \label{os1}
\begin{split}
P^{-}\tmatt{0}{n_1}{n_2}{0} &=  \frac{1}{\pt{n_1+n_2+1}\pt{n_1+n_2+2}\binom{n_1+n_2}{n_1}} \\
P^{-}\tmatt{s_1+1}{f_1-1}{s_2}{f_2} &= P^{-}\tmatt{s_1}{f_1}{s_2}{f_2} + \frac{P\tmatt{s_1+1}{f_1}{s_2}{f_2}}{n_1+1}  \\
P^{-} \tmatt{s_1}{f_1}{s_2-1}{f_2+1} &= P^{-} \tmatt{s_1}{f_1}{s_2}{f_2} + \frac{P\tmatt{s_1}{f_1}{s_2}{f_2+1}}{n_2+1} 
\end{split}
\end{equation}

\begin{figure}
	\includegraphics[width=\linewidth]{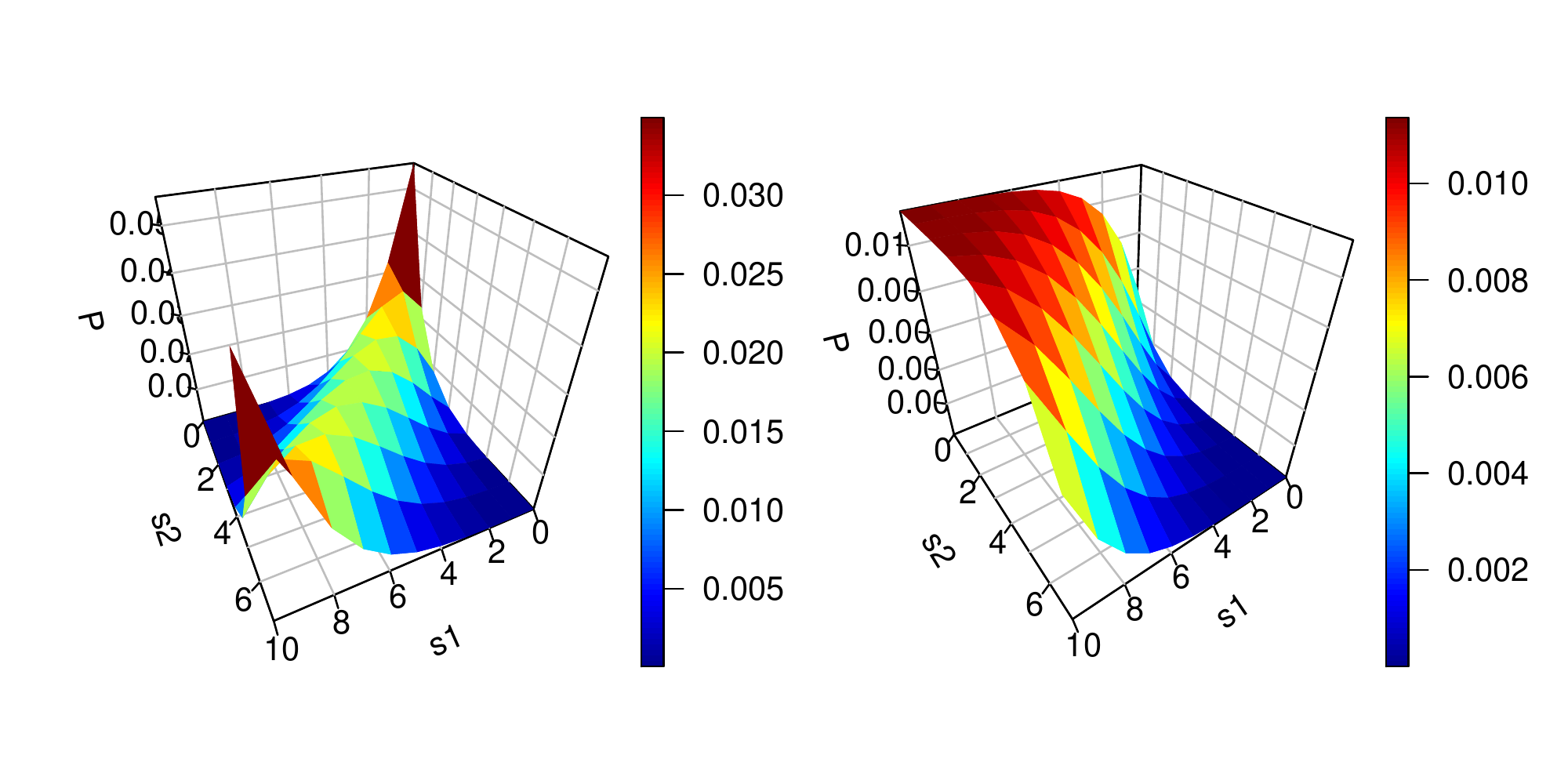}
	\caption{m-test probabilities ($P$) for a \emph{2 x 2} contingency table with column marginals $n_1 = s_1 + f_1 = 10$ and $n_2 = s_2 + f_2 = 7$. Left, two-sided test. Right, one-sided test.}
	\label{fig:fig1}
\end{figure}

Again, \eqref{os1} provides the basis for an algorithm to calculate the p-value under the one-sided null hypothesis with fixed column marginals (Figure \ref{fig:fig1}). In this case, the result must be normalized by multiplying by 2, since only half of the probability space for $q$ is integrated.

\section*{Implementation}

Using a generalized form of \eqref{eqsyst} explained in Proofs 6 and 7 of the Supplementary Material and \eqref{os1}, we coded the \texttt{mtest} package for \texttt{R}.This package uses \texttt{Rcpp} for speed and can calculate p-values for a \emph{2 x 2} table with column marginals of 400 in a few seconds on a standard laptop computer. It can also calculate two-sided p-values for arbitrary \emph{d x m} contingency tables. However, since the number of possible results with fixed column marginals escalates very quickly with additional outcomes, those marginals must be kept relatively low. As a reference, a p-value from a \emph{2 x 5} table and column marginals of 16 takes about 30 seconds to calculate in the same computer.

The \texttt{mtest} package also provides a Monte Carlo function to simulate results under the null hypothesis, both one-sided and two-sided. In addition to providing benchmarking capabilities for the package, this function can be used to find an approximate solution for complex contingency tables while limiting the number of calculation steps. Several examples are provided as \texttt{R notebooks} when downloading \texttt{mtest} to illustrate these features.

\section*{Test power}

\begin{figure}
	\includegraphics[width=\linewidth]{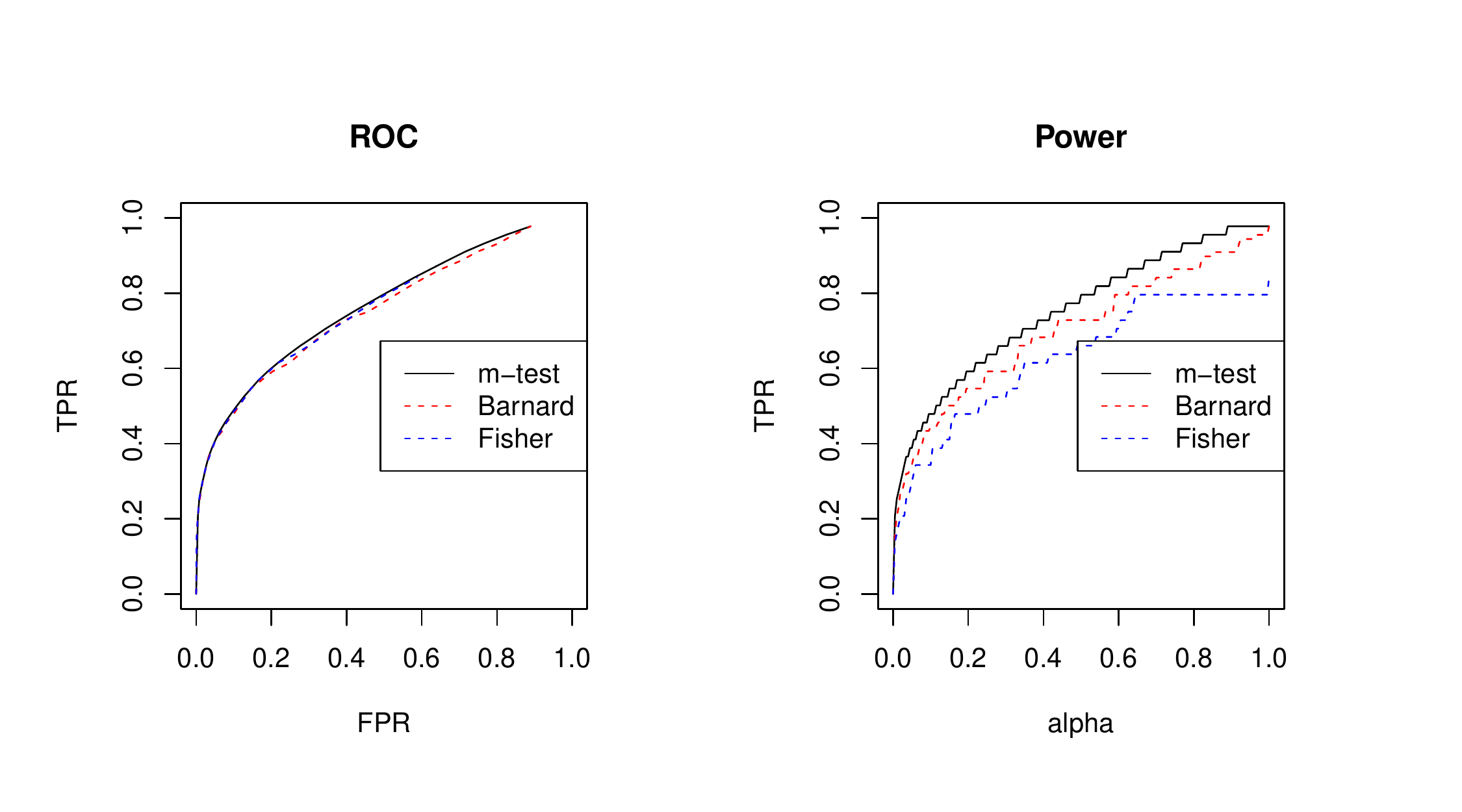}
	\caption{Receiver operation characteristic (ROC, left) and power (right) representations for m-test, Fisher's test and Barnard's test in simulated \emph{2 x 2} contingency tables with column marginals of 10 and 7. \emph{TPR}, true positive rate; \emph{FPR}, false positive rate; \emph{alpha}, significance level.}
	\label{fig:fig2}
\end{figure}

Using the Monte Carlo function, we simulated 200,000 results under the two-sided null hypothesis and 200,000 results under the alternative hypothesis. This allowed us to estimate the false positive and true positive rates at different significance levels. With these data, we characterized the discrimination capabilities and the power of the m-test compared to Fisher's test and Barnard's test (Figure \ref{fig:fig2}). In these benchmarks, the m-test consistently showed a higher statistical power than the other exact tests, while receiver operation characteristic (ROC) curves of all three tests were similar. Although power depends on the model chosen for the simulation and there are circumstances in which it may not hold, this result suggests that the m-test may offer a more powerful alternative to existing exact tests.

\section*{Methods}
The \texttt{mtest} package was coded in \texttt{R version 4.0.4} using the \texttt{Rcpp} \cite{rcpp} and \texttt{usethis} \cite{usethis} packages in \texttt{Rstudio 1.4.1106}. Multiple test results were benchmarked against the corresponding original definitions by numerical integration using \texttt{wxMaxima 16.04.2}. The package can be downloaded and installed from \url{https://github.com/vqf/mtest}. Figures were generated from \texttt{R notebooks} which are available at \url{https://github.com/vqf/mtest/tree/main/Figs}. Fisher's tests were run with the \texttt{fisher.test} function; Barnard's tests were calculated using the \texttt{Barnard} package \cite{pbarn}.
	
	\subsection*{Calculation of p-values}
	Both functions \texttt{m.test} (two-sided) and \texttt{os.m.test} (one-sided) perform the calculation of p-values following the corresponding equations in two steps. First, the probability of the problem result is calculated and used as an offset. Then, the probability of every possible result with the same column marginals is calculated. Each of those probabilities is added to the resulting p-value if it is not higher than the offset.
	
	In the one-sided test, according to equation \eqref{os1}, each probability is never lower than the previous value. This allows the calculation of p-values without visiting every possible result. This property can also be used to improve the calculation of two-sided p-values in the future, once the symmetries involved (apparent in Figure \ref{fig:fig1}) are characterized. On the other hand, column marginals higher than 100 can lead to underflow conditions, which prompt a warning from the package.
	
	\subsection*{Monte Carlo simulations}
	The m-test \texttt{mc} function simulates different \emph{d x m} contingency tables $N$ times by repeating two steps: probability assignment and trial. For the probability assignment of $d$ outcomes, $d-1$ pseudorandom numbers $\left\lbrace r_1, r_2, \ldots, r_{d-1}\right\rbrace \in \left[\left. 0, 1 \right)\right.$ are generated and sorted. For each trial, a pseudorandom number in $\left[\left. 0, 1 \right)\right.$ is generated. Outcome $o_i$ is selected if the trial number is in the $\left[\left. r_{i-1}, r_i \right)\right.$ interval, where $r_0 = 0$ and $r_d = 1$.
\section*{Acknowledgements}
M. A-V. is supported by the European Research Council. V. Q. is supported by the Spanish Ministerio de Ciencia, Innovaci\'{o}n y Universidades (RTI2018-096859-B-I00), including FEDER funding. The authors would like to thank Gonzalo Quesada and Dr. Magda Hamczyk for their advice, and Dr. Carlos L\'{o}pez-Ot\'{i}n for his work, his teachings and his dreams.
\newcommand{\gmat}{
	\left(\begin{smallmatrix}
		o_{1,1} & \cdots & o_{1,j} & \cdots & o_{1,m} \\
		o_{2,1} & \cdots & o_{2,j} & \cdots & o_{2,m} \\
		\vdots  & \ddots &  \vdots  & \ddots & \vdots  \\
		o_{d,1} & \cdots &  o_{d,j} & \cdots & o_{d,m} 
	\end{smallmatrix}\right)
}
\newcommand{\gmatpm}{
	\left(\begin{smallmatrix}
		o_{1,1} & \cdots & o_{1,j}+1 & \cdots & o_{1,m} \\
		o_{2,1} & \cdots & o_{2,j}-1 & \cdots & o_{2,m} \\
		\vdots  & \ddots &  \vdots  & \ddots & \vdots  \\
		o_{d,1} & \cdots &  o_{d,j} & \cdots & o_{d,m} 
	\end{smallmatrix}\right)
}
\newcommand{\gmatzero}{
	\left(\begin{smallmatrix}
		0 & \cdots & 0 & \cdots & 0 \\
		0 & \cdots & 0 & \cdots & 0 \\
		\vdots  & \ddots &  \vdots  & \ddots & \vdots  \\
		n_1 & \cdots &  n_j & \cdots & n_m
	\end{smallmatrix}\right)
}

\renewcommand{\theequation}{\thesection.\arabic{equation}}
\newcommand{\nextproof}{\addtocounter{section}{1}\setcounter{equation}{0}}

\section*{Supplementary Material}
\subsection*{Definitions}
	\begin{list}{defs}{}
		\item[Result:] Tabulation of multiple experiments with $n_j$ trials that yield different outcomes. In the case of a \emph{2 x m} contingency table, those outcomes can be labelled success ($s_j$) and failure ($f_j$), with $s_j+f_j=n_j$. The result will be expressed as a matrix with the numbers of successes and failures of each experiment, one experiment per column:
		\[
		\gtmat{s_1}{f_1}{s_j}{f_j}{s_m}{f_m}
		\]
		For the general \emph{d x m} contingency table, we label each outcome from $o_1$ to $o_d$:
		\[
		\gmat
		\]
		\item[Two-sided \emph{2 x m} null hypothesis:] The probability of success of each experiment ($\theta$) is the same. Therefore, the probability $P$ of obtaining the result is:
		\[
		P \gtmat{s_1}{f_1}{s_j}{f_j}{s_m}{f_m} = \binom{n_1}{s_1}\ldots\binom{n_j}{s_j}\ldots\binom{n_m}{s_m}\bp{\ssum{s_i}}{\ssum{f_i}}{}
		\]
		\item[One-sided null hypothesis:] For a \emph{2 x 2} contingency table, the probability of success in the first experiment ($\theta$) is higher than in the second experiment ($q$). Therefore, the probability of obtaining the result is:
		\[
		P^{-}\tmatt{s_1}{f_1}{s_2}{f_2} = \binom{n_1}{s_1} \binom{n_2}{s_2}\bp{s_1}{f_1}{\bm{s_2}{f_2}}
		\]
		\item[Two-sided with \emph{d x m} table:] Let $\theta_i$ be the probability of outcome $o_{i, j} \forall j \in \lbrace1, 2, \ldots m\rbrace$. By symmetry, since no outcome is privileged, any ordering of $\theta_i$ must yield the same result. Given that $\ssum{\theta_i} = 1$, there are $\pt{d-1}$ degrees of freedom and $\pt{d-1}!$ independent orderings of $\theta_i$. Taking $\theta_d$ as the dependent variable,
		\[
		\begin{multlined}
		P \gmat = \pt{d-1}!\prod_{j=1}^{m} \binom{n_j}{o_{1, j} o_{2, j} \ldots o_{d, j}}\int_{0}^{1}\theta_1^{\jsum{o_1, j}}\int_{1 - \theta_1}^{1}\theta_2^{\jsum{o_2, j}}\ldots \\
		\ldots\int_{1-\sum_{j=1}^{d-2}\theta_j}^{1}\theta_{d-1}^{\jsum{o_{d-1}, j}}\theta_d^{\jsum{o_{d, j}}}d\theta_{d-1}\ldots d\theta_2 d\theta_1
		\end{multlined}
		\]
		Let $l_i = 1 - \sum_{j=1}^{i}\theta_j$. Therefore, $\theta_1 = \pt{1 -l_1}$, $l_d = 0$, $\theta_i = l_{i-1} - l_{i}$ and $d\theta_i = -dl_i$ (as $l_{i-1}$ does not include $\theta_i$).
		\[
		\begin{multlined}
		P \gmat = \pt{d-1}!\prod_{j=1}^{m} \binom{n_j}{o_{1, j} o_{2, j} \ldots o_{d, j}}\int_{0}^{1}\pt{1-l_1}^{\jsum{o_1, j}}\ldots \\
		\ldots\int_{0}^{l_1}\pt{l_1-l_2}^{\jsum{o_2, j}}\int_{0}^{l_{d-2}}\pt{l_{d-2}-l_{d-1}}^{\jsum{o_{d-1}, j}}l_{d-1}^{\jsum{o_{d, j}}}dl_{d-1}\ldots dl_2 dl_1
		\end{multlined}
		\]
	\end{list}
	\subsection*{Proofs}
	\nextproof
	\begin{proof}[Proof 1]
		$P \gtmat{0}{n_1}{0}{n_j}{0}{n_m} = \frac{1}{\tsum}$
		\begin{align}
		P \gtmat{0}{n_1}{0}{n_j}{0}{n_m} &= \cancelto{1}{\binom{n_1}{0}\ldots\binom{n_j}{0}\ldots\binom{n_m}{0}}\bp{0}{\sum_{i=1}^{m}{n_i}}{} = \\
		&= -\frac{1}{\tsum}\left[\left(1-\theta\right)^{\tsum}\right]_0^1 = \frac{1}{\tsum}\qedhere
		\end{align}
	\end{proof}
	\nextproof
	
	\begin{proof}[Proof 2]
		$P \gtmat{s_1}{f_1}{s_j+1}{f_j - 1}{s_m}{f_m} = \frac{f_j\pt{1+\ssum{s_i}}}{\pt{s_j+1}\ssum{f_i}} P \gtmat{s_1}{f_1}{s_j}{f_j}{s_m}{f_m}$
		\begin{align}
		P \gtmat{s_1}{f_1}{s_j+1}{f_j - 1}{s_m}{f_m} &= \binom{n_1}{s_1}\ldots\binom{n_j}{s_j+1}\ldots\binom{n_m}{f_m} \bp{1+\ssum{s_i}}{-1+\ssum{f_i}}{} \label{note1} \\ 
		\intertext{Integrating by parts with:
			\begin{tabular}{|l|l|}
			$u = \theta^{1+\ssum{s_i}}$ & $dv = \pt{1-\theta}^{\ssum{f_i}-1} d\theta$ \\
			$du = \pt{1+\ssum{s_i}}\theta^{\ssum{s_i}}d\theta$ & $v = -\frac{\pt{1-\theta}^{\ssum{f_i}}}{\ssum{f_i}} $ \\
			\end{tabular}
		}
		&= \begin{multlined}[t]
		\binom{n_1}{s_1}\ldots\frac{f_j}{s_j+1}\binom{n_j}{s_j}\ldots\binom{n_m}{f_m} \frac{1}{\ssum{f_i}}\cancelto{0}{\left[\theta^{1+\ssum{s_i}}\pt{1-\theta}^{\ssum{f_i}}\right]_0^1}+\\
		+\binom{n_1}{s_1}\ldots\frac{f_j}{s_j+1}\binom{n_j}{s_j}\ldots\binom{n_m}{f_m} \frac{1+\ssum{s_i}}{\ssum{f_i}}\bp{\ssum{s_i}}{\ssum{f_i}}{}
		\end{multlined} \\
		&= \frac{f_j\pt{1+\ssum{s_i}}}{\pt{s_j+1}\ssum{f_i}}\binom{n_1}{s_1}\ldots\binom{n_j}{s_j}\ldots\binom{n_m}{f_m}\bp{\ssum{s_i}}{\ssum{f_i}}{} \\
		&= \frac{f_j\left(1 + \sum_{i=1}^{m}s_i\right)}{\left(1 + s_i\right)\ssum{f_j}} P \gtmat{s_1}{f_1}{s_j}{f_j}{s_m}{f_m}\qedhere
		\end{align}
		
	\end{proof}
	\nextproof
	\begin{proof}[Proof 3]	
		$P \gmatzero = \frac{\pt{d-1}!}{1 + \ssum{n_i}}$
		\begin{align}
		P \gmatzero &= \pt{d-1}!\prod_{j-1}^{m}\binom{n_j}{n_j}\int_{0}^{1}\int_{0}^{l_1}\ldots\int_{0}^{l_{d-2}}l_{d-1}^{\ssum{n_i}}dl_{d-1}\ldots dl_2 dl_1 \\
		&= \frac{\pt{d-1}!}{1 + \ssum{n_i}}\cancelto{1}{\ct{\ct{\ldots\ct{l_{d-1}^{1 + \ssum{n_i}}}_{0}^{l_{d-2}}\ldots}_{0}^{l_{1}}}_{0}^{1}}\qedhere
		\end{align}
		
	\end{proof}
	\nextproof
	\begin{proof}[Proof 4]	
		$P \gmatpm = \frac{o_{2, j}\pt{1+\ssum{o_{1}, i}}}{\pt{o_{1, j}+1}\ssum{o_{2, i}}} P \gmat$
		
		\text{Without loss of generality we can consider} 
		\begin{align}
		P \gmatpm \MoveEqLeft[5]= \begin{multlined}[t]\pt{d-1}!\binom{n_j}{\pt{o_{1, j}+1} \pt{o_{2, j}-1} \ldots o_{d, j}} \\
		\prod_{\substack{k=1 \\ k \ne j}}^{m} \binom{n_k}{o_{1, k} o_{2, k} \ldots o_{d, k}}\int_{0}^{1}\pt{1-l_1}^{\ssum{o_3, i}}\int_{0}^{l_1}\pt{l_1-l_2}^{\ssum{o_4, i}}
		\ldots \\
		\ldots\int_{0}^{l_{d-2}}\pt{l_{d-2}-l_{d-1}}^{\pt{1+\ssum{o_{1}, i}}}l_{d-1}^{\pt{-1+\ssum{o_{2, i}}}}dl_{d-1}\ldots dl_2 dl_1
		\end{multlined} \\
		\intertext{Solving the innermost integral by parts with }
		\intertext{
			\begin{tabular}{|l|l|}
			$u = \pt{l_{d-2}-l_{d-1}}^{\pt{1+\ssum{o_{1}, i}}} $ & $dv=l_{d-1}^{\pt{-1+\ssum{o_{2, i}}}}dl_{d-1}$ \\[1em]
			$du = -\pt{1+\ssum{o_{1}, i}}\pt{l_{d-2}-l_{d-1}}^{\ssum{o_{1}, i}}dl_{d-1}$ & $v = \frac{l_{d-1}^{\ssum{o_{2, i}}}}{\ssum{o_{2, i}}}$ 			
			\end{tabular}}
		\intertext{
			\begin{multlined}[t]
			\int_{0}^{l_{d-2}}\pt{l_{d-2}-l_{d-1}}^{\pt{1+\ssum{o_{1}, i}}}l_{d-1}^{\pt{-1+\ssum{o_{2, i}}}}dl_{d-1} = \cancelto{0}{\ct{\pt{l_{d-2}-l_{d-1}}^{\pt{1+\ssum{o_{1}, i}}} \frac{l_{d-1}^{\ssum{o_{2, i}}}}{\ssum{o_{2, i}}}}_{l_{d-1} = 0}^{l_{d-1} = l_{d-2}}} +\\ 
			\\
			+ \frac{1+\ssum{o_{1}, i}}{\ssum{o_{2, i}}}\int_{0}^{l_{d-2}}\pt{l_{d-2}-l_{d-1}}^{\ssum{o_{1}, i}}l_{d-1}^{\ssum{o_{2, i}}} dl_{d-1}
			\end{multlined}} 
		P \gmatpm &= \begin{multlined}[t]\pt{d-1}!\frac{o_{2, j}}{o_{1, j}+1}\prod_{k=1}^{m} \binom{n_k}{o_{1, k} o_{2, k} \ldots o_{d, k}} \\
		\frac{1+\ssum{o_{1}, i}}{\ssum{o_{2, i}}}\int_{0}^{1}\pt{1-l_1}^{\ssum{o_3, i}}\int_{0}^{l_1}\pt{l_1-l_2}^{\ssum{o_4, i}}
		\ldots \\
		\ldots\int_{0}^{l_{d-2}}\pt{l_{d-2}-l_{d-1}}^{\ssum{o_{1}, i}}l_{d-1}^{\ssum{o_{2, i}}}dl_{d-1}\ldots dl_2 dl_1 \label{multinom} 
		\end{multlined} \\
		&= \frac{o_{2, j}\pt{1+\ssum{o_{1}, i}}}{\pt{o_{1, j}+1}\ssum{o_{2, i}}} P \gmat\qedhere
		\end{align}
		
	\end{proof}
	
	\nextproof
	\begin{proof}[Proof 5]
		$P^{-}\tmatt{0}{n_1}{n_2}{0} =  \frac{1}{\pt{n_1+n_2+1}\pt{n_1+n_2+2}\binom{n_1+n_2}{n_1}}$
		\begin{align}
		P^{-}\tmatt{0}{n_1}{n_2}{0} &= \binom{n_1}{0} \binom{n_2}{n_2}\bp{0}{n_1}{\bm{n_2}{0}} \\
		\bm{n_2}{0} &= \frac{1}{n_2+1}\left[q^{n_2+1}\right]_{q=0}^{\theta} = \frac{\theta^{n_2+1}}{n_2+1} \\
		P^{-}\tmatt{0}{n_1}{n_2}{0} &= \frac{1}{n_2+1}\bp{n_2+1}{n_1}{} \label{euler} \\
		&= \frac{1}{\pt{n_2+1}\pt{n_1+n_2+2}\binom{n_1+n_2+1}{n_1}} = \frac{1}{\pt{n_1+n_2+1}\pt{n_1+n_2+2}\binom{n_1+n_2}{n_1}} \qedhere
		\end{align}
	\end{proof}

	\nextproof
	\begin{proof}[Proof 6]
		$P^{-}\tmatt{s_1+1}{f_1-1}{s_2}{f_2} = P^{-}\tmatt{s_1}{f_1}{s_2}{f_2} + \frac{P\tmatt{s_1+1}{f_1}{s_2}{f_2}}{n_1+1}$
		\begin{align}
		P^{-}\tmatt{s_1+1}{f_1-1}{s_2}{f_2} &= \binom{n_1}{s_1+1} \binom{n_2}{s_2}\bp{s_1+1}{f_1-1}{\bm{s_2}{f_2}} \\
		\intertext{By the fundamental theorem of calculus,}
		&\frac{d\pt{\bm{s_2}{f_2}}}{d\theta} = \theta^{s_2}\pt{1-\theta}^{f_2}
		\intertext{Integrating by parts with:} 
		\intertext{	
		\begin{tabular}{|l|l|}
			$u = \theta^{s_1+1}\bm{s_2}{f_2}$ & $dv = \pt{1-\theta}^{f_1-1} d\theta$ \\
			$du = \pt{s_1+1}\theta^{s_1}\bm{s_2}{f_2} + \theta^{s_1+s_2+1}\pt{1-\theta}^{f_2}dp$ & $v = -\frac{1}{f_1} \pt{1-\theta}^{f_1}$ \\
			\end{tabular}}
		P^{-}\tmatt{s_1+1}{f_1-1}{s_2}{f_2} &= \begin{multlined}[t]
		-\binom{n_1}{s_1 + 1} \binom{n_2}{s_2}\cancelto{0}{\ct{\theta^{s_1+1}\bm{s_2}{f_2}\frac{1}{f_1} \pt{1-\theta}^{f_1}}_{\theta=0}^{1}} + \\
		+ \binom{n_1}{s_1 + 1} \binom{n_2}{s_2}\frac{s_1+1}{f_1}\bp{s_1}{f_1}{\bm{s_2}{f_2}}+ \\
		+ \binom{n_1}{s_1 + 1}\binom{n_2}{s_2}\frac{1}{f_1}\bp{s_1+s_2+1}{f_1+f_2}{} 
		\end{multlined} \label{note2} \\
		&= \begin{multlined}[t]
		\binom{n_1}{s_1}\binom{n_2}{s_2}\bp{s_1}{f_1}{\bm{s_2}{f_2}} + \\
		+\binom{n_1+1}{s_1 +  1}\binom{n_2}{s_2}\frac{1}{n_1+1}\bp{s_1+s_2+1}{f_1+f_2}{}
		\end{multlined} \\
		&=P^{-}\tmatt{s_1}{f_1}{s_2}{f_2} + \frac{P\tmatt{s_1+1}{f_1}{s_2}{f_2}}{n_1+1}\qedhere
		\end{align}
	\end{proof}
	
	\nextproof
	\begin{proof}[Proof 7]
		$P^{-} \tmatt{s_1}{f_1}{s_2-1}{f_2+1} = P^{-} \tmatt{s_1}{f_1}{s_2}{f_2} + \frac{P\tmatt{s_1}{f_1}{s_2}{f_2+1}}{n_2+1}$
		\begin{align}
		P^{-}\tmatt{s_1}{f_1}{s_2-1}{f_2+1} &= \binom{n_1}{s_1} \binom{n_2}{s_2-1}\bp{s_1}{f_1}{\bm{s_2-1}{f_2+1}} \\
		\intertext{Solving the second integral by parts with:
			\begin{tabular}{|l|l|}
			$u = \pt{1-q}^{f_2+1}$ & $dv = q^{s_2-1} dq$ \\
			$du = -\pt{f_2+1}\pt{1-q}^{f_2}dq$ & $v = \frac{1}{s_2} q^{s_2}$
			\end{tabular}
		} \\
		&= \begin{multlined}[t]
		\binom{n_1}{s_1} \binom{n_2}{s_2-1} \\
		\bp{s_1}{f_1}{\ct{\ct{\frac{1}{s_2}q^{s_2}\pt{1-q}^{f_2+1}}_{q = 0}^{\theta}+\frac{f_2+1}{s_2}\bm{s_2}{f_2}}} 
		\end{multlined} \\
		&= \begin{multlined}[t]
		\frac{f_2+1}{s_2}\binom{n_1}{s_1} \binom{n_2}{s_2-1}\bp{s_1}{f_1}{\bm{s_2}{f_2}} + \\
		+ \binom{n_1}{s_1} \binom{n_2}{s_2-1}\frac{1}{s_2}\bp{s_1+s_2}{f_1+f_2+1}{} \end{multlined} \label{note3} \\
		&= \begin{multlined}[t]
		\binom{n_1}{s_1} \binom{n_2}{s_2}\bp{s_1}{f_1}{\bm{s_2}{f_2}} + \\
		+ \binom{n_1}{s_1} \binom{n_2+1}{s_2}\frac{1}{n_2+1}\bp{s_1+s_2}{f_1+f_2+1}{} \end{multlined} \\
		&= P^{-}\tmatt{s_1}{f_1}{s_2}{f_2} + \frac{P\tmatt{s_1}{f_1}{s_2}{f_2+1}}{n_2+1}\qedhere
		\end{align}
	\end{proof}

	\subsection*{Notes}
	\begin{list}{Notes}{}
		\item[\ref{note1}] \begin{align*}\binom{n_j}{s_j+1} = \frac{\pt{n_j}!}{\pt{s_j+1}!\pt{f_j-1}!} =  \frac{\pt{n_j}!}{\pt{s_j+1}s_j!\frac{f_j!}{f_j}} = \frac{f_j}{s_j+1}\binom{n_j}{s_j} \end{align*}
		\item[\ref{multinom}]
		\begin{align*}
		\binom{n_j}{\pt{o_{1, j}+1} \pt{o_{2, j}-1} \ldots o_{d, j}} &= \frac{n_j!}{\pt{o_{1, j}+1}!\pt{o_{2, j}-1}!\ldots o_{d, j}!} = \frac{n_j!}{\pt{o_{1, j}}!\pt{o_{1, j}+1}\frac{\pt{o_{2, j}}!}{o_{2, j}}\ldots o_{d, j}!} \\
		&= \frac{o_{2, j}}{o_{1, j}+1}\binom{n_j}{o_{1, j} o_{2, j} \ldots o_{d, j}}
		\end{align*}
		\item[\ref{euler}] Euler integral of the first kind
		\item[\ref{note2}] \begin{align*}
		\binom{n_1}{s_1+1}\frac{s_1+1}{f_1} &= \frac{n_1!\pt{s_1+1}}{\pt{s_1+1}!\pt{f_1-1}!f_1} = \frac{n_1!}{s_1!f_1!} = \binom{n_1}{s_1} \\
		\binom{n_1}{s_1+1}\frac{1}{f_1} &= \frac{n_1!}{\pt{s_1+1}!\pt{f_1-1}!f_1} = \frac{1}{n_1+1}\frac{\pt{n_1+1}!}{\pt{s_1+1}!f_1!} = \frac{1}{n_1+1}\binom{n_1+1}{s_1+1}
		\end{align*}
		\item[\ref{note3}] 
		\begin{align*}
		\binom{n_2}{s_2-1}\frac{f_2+1}{s_2} &= \frac{n_2!\pt{f_2+1}}{\pt{s_2-1}!\pt{f_2+1}!s_2} = \frac{n_2!}{s_2!f_2!} = \binom{n_2}{s_2} \\
		\binom{n_2}{s_2-1}\frac{1}{s_2} &= \frac{n_2!}{\pt{s_2-1}!s_2\pt{f_2 + 1}!} = \frac{1}{n_2+1}\frac{\pt{n_2+1}!}{s_2!\pt{f_1+1}!} = \frac{1}{n_2+1}\binom{n_2+1}{s_2}
		\end{align*}
		
	\end{list}

\bibliographystyle{unsrt}  
\bibliography{mtest}  

\begin{thebibliography}{1}

\bibitem{recommended}
S.~Lydersen, M.~W. Fagerland, and P.~Laake.
\newblock Recommended tests for association in 2 x 2 tables.
\newblock {\em Stat Med}, 28:1159--75, 2009.

\bibitem{Fisher}
R.~A. Fisher.
\newblock The logic of inductive inference.
\newblock {\em Journal of the Royal Statistical Society}, 98(1):39--54, 1935.

\bibitem{Design}
R.~A. Fisher.
\newblock {\em The design of experiments}.
\newblock Oliver \& Boyd, 1935.

\bibitem{Barnard}
G.~A. Barnard.
\newblock A new test for 2 × 2 tables.
\newblock {\em Nature}, 156:177, 1945.

\bibitem{fview}
Enrico Ripamonti, Chris Lloyd, and Piero Quatto.
\newblock {Contemporary Frequentist Views of the $2\times2$ Binomial Trial}.
\newblock {\em Statistical Science}, 32(4):600 -- 615, 2017.

\bibitem{rcpp}
Dirk Eddelbuettel and James~Joseph Balamuta.
\newblock Extending r with c++: A brief introduction to rcpp.
\newblock {\em The American Statistician}, 72(1):28--36, 2018.

\bibitem{usethis}
Hadley Wickham and Jennifer Bryan.
\newblock {\em usethis: Automate Package and Project Setup}, 2021.
\newblock R package version 2.0.1.

\bibitem{pbarn}
Kamil Erguler.
\newblock {\em Barnard: Barnard's Unconditional Test}, 2016.
\newblock R package version 1.8.

\end{thebibliography}

\end{document}